# Evidences of Superconductivity without Direct Relation to Valence Electron Density in Metal Doped Y123 Cuprates[*]

WANG Xiao-Xia, YU Guo-Ru, LI Jiang-Hui, LI Ping-Lin

*School of Physics and Engineering, Zhengzhou University, Zhengzhou 450052, PRC*

**Abstract**  We investigated systemically $YBa_2Cu_{3-x}(Fe, Co, Al, Zn)_xO_{7-\delta}$ ($x=0.0 \sim 0.5$) cuprates through the XRD, the test of superconductivity, oxygen contents, positron annihilation technology and simulated calculations. The experimental results and theoretic calculations support such a conclusion that the $T_c$ does not depend directly on the density of valence electron in the samples. Moreover, whether or not it may imply some correlations between Fe-oxide superconductors and cuprates that Fe doped Y123 samples show some unique characteristics.

*Keywords:* metal ion doped Y123, simulated calculations, positron annihilation, cluster effect, valence electron density

*PACS numbers*: 47.54.De, 47.54.Bd, 68.65.-k

[*] **This work is supported by The Natural Science Foundation of China (NO. 10647145).**
LI Ping-Lin Email: lipinglin@zzu.edu.cn

## 1. Introduction

Due to iron arsenide superconductors were discovered [1, 2] and are comparable with the high-$T_c$ cuprates in the many properties [3, 4], the later becomes again into a highlight in the superconducting researches [5-11]. However, the microcosmic mechanism is challenging harshly to all theories, even as Zaanens statements: "The high-$T_c$ superconductivity is on the list of the most profound physics problems......"[12]. In order to search and comprehend such problems, scholars suggested many new theories and experimental technologies[13-17], for example, elemental substitution that play an important role in the research of high-$T_c$ cuprates[18, 19], even iron arsenides, until today, namely, the copper or ferrum is replaced by magnetic or nonmagnetic elements[1-4]. According to the traditional theory of magnetic break-pair, the nonmagnetic ions doped into superconductors will suppress less the superconductivity than magnetic ions. However, $Zn^{2+}$ doped in Y123, a kind of nonmagnetic ion, suppresses more forcefully the superconductivity than other magnetic ions, like Fe and Co. Further, the new iron arsenide superconductors completely overthrown the traditional theory that the magnetism depresses the superconductivity in the samples. As the representative of magnetic and nonmagnetic ion, Fe, Co, Al and Zn doped in Y123, they show some special characteristics, for instance, Zn doping suppresses most forcefully the superconductivity in four kinds of ions; nevertheless, as the same nonmagnetic ion as Zn, Al doping is opposite to Zn doping in suppression on the superconductivity; furthermore Fe doped samples do not show any impure phases until $x = 0.50$ in the experiments, such a result may imply some possibility in the existence of Fe-oxide (iron arsenide) superconductors[1-4]. Many queries are now still unclear; therefore, further investigations of theories and experiments will be necessary in the metal doped Y123 cuprates.

In order to realize the distribution of doped ions, here will calculate the total defect energy and average binding energy inside a cluster on Islam's method [20], simulated results reveal the possibility of cluster effect. When the doped concentration increases, the doped ions combine into different clusters on $CuO_2$ planes and in CuO chains. Afterwards, we use positron annihilation technique (PAT) probes the change of electron structure as four kinds of metal ions substitute for Cu in Y123. PAT plays an important role in the investigation of such condensed matters as semiconductors, metal materials and high-$T_c$ cuprates [21-25]. Some positron studies exhibit the properties of normal and superconducting states, such as the Fermi surface, O-T transition and the carrier concentration [26, 27]. In recent years, Jean and Li et al. reported many important results from positron experiments and theories concerning high-$T_c$ superconducting mechanism [28-35]. In the present article, we will report the systematical investigation concerning $YBa_2Cu_{3-x}$(Fe, Co, Al)$_x$O$_{7-\delta}$ (x=0.0~0.5) samples, for the sake of comparison with Zn doped cuprates, here mentions some correlated contents in Reference 31(LI *et. al.* 2008). The theoretic calculations and experimental results of PAT and oxygen content support such a conclusion that the superconductivity have no direct relation to the density of valence electron in the samples. Simultaneously, whether or not it may imply some correlations between the Fe-oxide superconductors and cuprates that Fe doped cuprates show some special characteristics.

## 2. Experiments and Simulated calculations

Samples of YBa$_2$Cu$_{3-x}$(Fe, Co, Al)$_x$O$_{7-\delta}$ ($x$=0~0.5) were prepared by the same method in Ref. 28-32. Due to the sensitivity of the PAT experiments, all samples were sintered in the same conditions in order to reduce the dispersion of the experimental results. The superconducting transition temperature $T_c$ was measured by the standard *DC* four-probe method with a voltage resolution of *10$^{-7}$ V* (*HP3457A*). The crystal structures of samples were analyzed by powder X-ray diffraction *(Cu-K$_\alpha$)* using the *D/max-BX-ray* diffractometer. The positron lifetime spectra were measured by the *ORTEC-100U* fast-fast coincidence lifetime spectrometer. Two pieces of identical samples (*Φ13×3mm*) were sandwiched together with a 10-*μC22 Na* positron source deposited on a thin Mylar foil (about *1.2mg/cm$^2$* thickness). We apply Pilot-U plastic flicker sensor，which was measured by $^{60}$Co and showed the excellent time resolution over 220ps. Each spectrum contains more than *1×10$^6$* counts to guarantee the sufficient statistic precision. After subtracting background and source contributions, the lifetime spectra were fitted with two-lifetime components by *POSITRON-FIT-EXTENDED* program with the best fit ($\chi^2$ = 1.0~1.1 ). The positron lifetime spectra of all samples were measured at the identical environment temperature (283±1*K*)，and the results were repeatable. Table 2-4 list the main experimental results of Fe, Co and Al doped cuprates, including $T_c$, lattice parameters, positron annihilation parameters etc. While the experimental results of Zn doped systems are here omitted, if necessary, please read LI *et. al.* paper (2008) in Ref. 31.

In order to understand the characteristics of cluster structure, here performs simulated calculations on the energy minimization principle and framework of Born model, in which the effective pairwise potentials represent the interatomic forces in the following form [20]:

$$\Phi_{ij} = \frac{Z_i Z_j e^2}{4\pi\varepsilon_0 r_{ij}} + A_{ij}\exp\{-\frac{r_{ij}}{\rho_{ij}}\} - \frac{C_{ij}}{r_{ij}^6}, \qquad (1)$$

Where $\Phi_{ij}$ is the effective potentials between ion *i* and *j*; $Z_i$ and $Z_j$ are ion valences; $r_{ij}$ is the distance between ion *i* and *j*; $A_{ij}$, $C_{ij}$ and $\rho_{ij}$ are the relative character constants. The first term is the long-range Coulomb interaction; the remaining terms represent the short-range interaction and the shielded revision. Because charge equilibrium determines the ion coordination characteristics in clusters, the clusters should be neutral in electricity. The neutral clusters may show several structures, two of which are illustrated as (a) and (b) on CuO chains in Ref 28, 29. Here (a) as Hexamer denotes six doped ions with two common-lateral squares, and (b) as Double-Square does seven doped ions with two common-apical squares. In the case of substitution for Cu, Fe, Co and Al ions can be regarded as defects. Similarly, an additional or missing oxygen ion is yet a defect. In the Cu-O chains, the valences of Cu and doped M (Fe, Co, Al) ions are +2 and +3, respectively. An $O^{2-}$ requires two valence electrons, but a solitary dispersive $M^{3+}$ ion can supply only one valence electron, *i.e.*, it cannot capture one oxygen ion for charge equilibrium, which can only be reached when the cluster effect occurs and the extra oxygen ions enter a cluster. The simplest neutral cluster is a dimer consisted by an oxygen ion $O^{2-}$ and two neighboring $M^{3+}$ ions, namely such a form $M^{3+}-O^{2-}-M^{3+}$. The binding energy of the dimer cluster is:

$$E_{bind} = E_{b1}(2M^{3+} \to 2Cu^{2+} + O^{2-}) - 2E_{b2}(M^{3+} \to Cu^{2+}) - E_{b3}(O^{2-}) \qquad (2)$$

Where $M^{3+}$ represents the Fe, Co and Al ions, and the first term $E_{b1}$ is the total defect energy of a cluster, while $E_{b2}$ and $E_{b3}$ are the energies of an isolated $M^{3+}$ and an isolated oxygen ion $O^{2-}$, respectively. These $E_{b1}$, $E_{b2}$ and $E_{b3}$ are all obtained by the formula $E = \sum \Phi_{ij}$. Every doped ion has the average binding energy $E_{mean} = E_{bind} / N$, where $N$ is the number of doped ions in the cluster. Other clusters may likewise be described and calculated. The calculation results are all listed in Table 1, in which the negative value of the binding energy indicates the system is bound. As well known, the larger average binding energy is, the more stable a cluster combines. Consequently, the doped ions should prefer to form the cluster with largest average binding energy. It can be concluded from the results listed in Table 1 that doped $M^{3+}$ ions prefer to form Hexamer cluster of six ions, while for Zn doping Double Square (a common point, seven ions) clusters have the largest probability. As for the larger cluster can be composed by some small clusters, for example, the cluster of eight ions are by double four ions and so on.

## 3. Results and discussions

### 3.1 XRD results and cuprate superconductivity

XRD results show undoped and low-doped Y-123 samples have the well single phase. As

**Table 1. The $Fe^{3+}$, $Co^{3+}$ and $Al^{3+}$ mean binding energy of per ion in the cluster.**

| Cluster | $Fe^{3+}$-$E_m$ | $Co^{3+}$-$E_m$ | $Al^{3+}$-$E_m$ |
|---------|-----------------|-----------------|-----------------|
| D+O     | −2.90           | −2.80           | −2.73           |
| TL+2O   | −2.57           | −2.44           | −2.25           |
| TS+2O   | −2.89           | −2.82           | −2.71           |
| TZ+2O   | −2.16           | −2.15           | −2.10           |
| H+3O    | −2.93           |                 | −2.83           |
| H+4O    | −3.19           | −3.03           | −2.85           |
| DS+4O   | −3.02           | −2.91           | −2.81           |

**Note**: Here the negative value indicates the binding energy. D+O=Dimmer $\{2M^{3+}\rightarrow 2Cu^{2+}+O^{2-}\}$, "+O" represents an $O^{2-}$ ion is attracted into the cluster; TL+2O=Tetramer$\{4\ M^{3+} \rightarrow 4Cu^{2+}+2O^{2-}\}$Linear; TS+2O=Tetramer$\{4\ M^{3+} \rightarrow 4Cu^{2+}+2O^{2-}\}$Square; TZ+2O=Tetramer $\{4\ M^{3+} \rightarrow 4Cu^{2+}+2O^{2-}\}$ Zigzag; H+4O=Hexamer $\{6M^{3+}\rightarrow 6Cu^{2+}+4O^{2-}\}$; DS+4O=Double Square $\{7M^{3+}\rightarrow 7Cu^{2+}+4O^{2-}\}$.

doped concentration $x$ increase, samples show slight impure phases at $x = 0.12$, $x = 0.20$ and $x=0.25$ for Zn, Al and Co doping, respectively. However, Fe doped samples do not show any impure phases until $x = 0.50$. Here in-order four $x$ values correspond with the mean binding energy $E_m$ of Fe, Co, Al and Zn every cluster in Table 1 and Ref 26, namely, $x$ (Fe) > $x$ (Co) > $x$ (Al) > $x$ (Zn) has the same relation to $E_m$(Fe) > $E_m$(Co) > $E_m$(Al) > $E_m$(Zn). Such a result reveals that the small theoretic $E_m$ samples occur easily the impure phases in the experiments, so our theoretic calculations are bale to reflect the experimental data.

On the other hand, as above statements Fe doped samples do not show any impure phases until $x = 0.50$, which reveals that Fe ions play almost a same role as Cu in the crystal texture of high-$T_c$ cuprates, whether or not such a result may imply some correlations between the Fe-oxide superconductors[1-4] and cuprates, perhaps, this issue could be replied through the further investigations concerning both high-$T_c$ superconductors.

We achieved the lattice parameters were calculated by the least-square method with powder

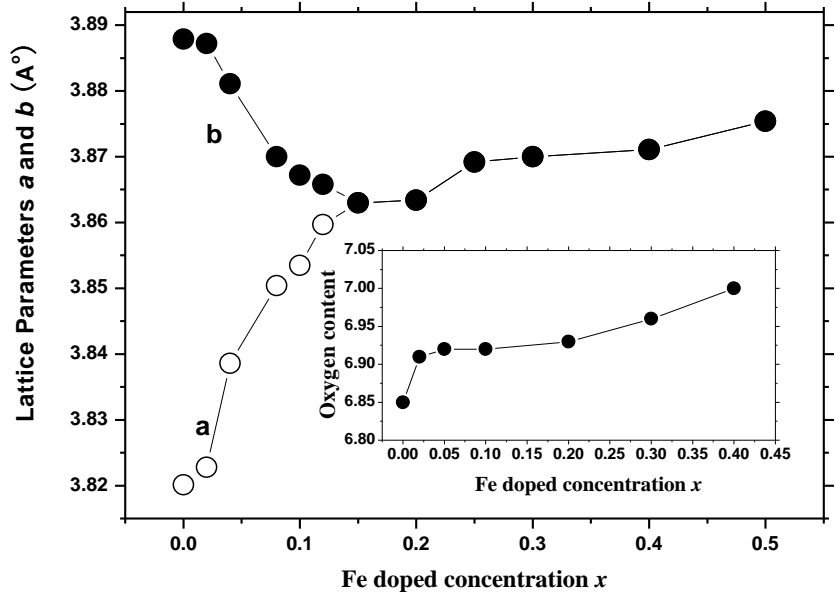

Figure 1. Lattice parameter *a* and *b* variation with Fe doped concentration *x* in $YBa_2Cu_{3-x}Fe_xO_{7-\delta}$ systems. And the inlet picture indicates the oxygen content variation with *x*.

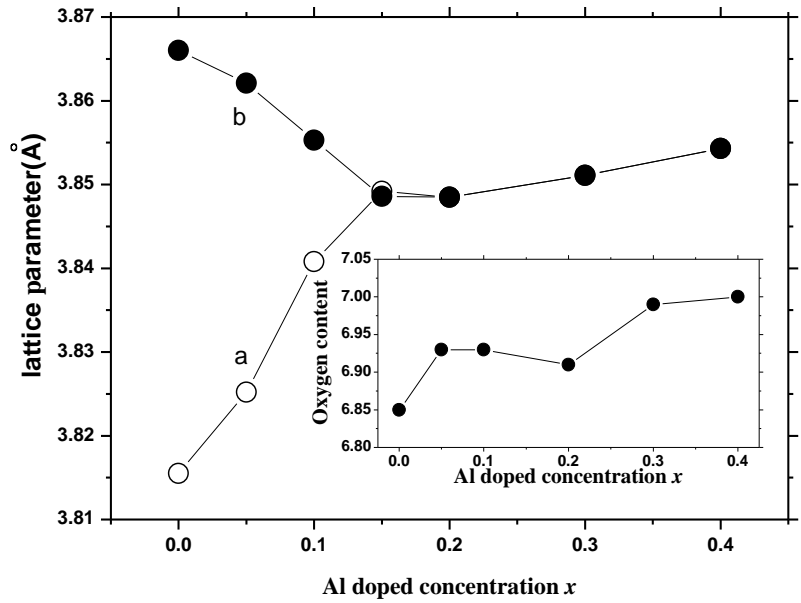

Figure 2  Lattice parameter *a* and *b* variation with Al doped concentration *x* in $YBa_2Cu_{3-x}Al_xO_{7-\delta}$ systems. And the inlet picture indicates the oxygen content variation with *x* (Ref. 41).

XRD data, and the results of Fe and Al doped cuprates are shown in Figure 1 and Figure 2. Because Co doped samples have the similar data to Fe and Al doped, the lattice parameters of Co doping are not illustrated. As for the results of Zn doped cuprates, if necessary, please read LI *et. al.* paper (2008) in Ref. 31. When Fe and Al concentration $x$ increase, as showed in Figure 1 and Figure 2, the O-T transition appears near x=0.15, nevertheless, for Co doped cuprates, the O-T transition appears near x=0.12 as listed in Table 3. For Zn doped samples, the lattice parameter $a$ and $b$ increase slightly with $x$ increasing, this is because the $Zn^{2+}$ radius (0.74 Å) is larger than the $Cu^{2+}$ radius (0.72 Å), but the O-T transition does not appear in $x$=0.0~0.4.

Generally, O-T transition results from the change of oxygen content in Cu-O chains. Thus the variation of lattice parameters $a$ and $b$ indicates that Fe, Co and Al ions enter mainly Cu(1) sites, which is consistent with Bringley and Hoffmann *et al* experiments [36, 37]. For the same reason, the variation of lattice parameters without O-T transition shows that Zn ions enter mainly Cu(2) sites on $CuO_2$ planes [38-40], where the superconductivity occurs, this implies that the superconductivity suppressed by Zn doping should be stronger than by Fe, Co and Al doping. In our experiments, the $T_c$ is 92K for undoped Y123 compound as listed in Table 2-4. For Fe doping, during $x$=0~0.04, the $T_c$ falls to 84.0K; at $x$=0.10, the $T_c$ is 75.3K; at $x$=0.15, the $T_c$ is 55.5K; at $x$=0.20, the $T_c$ is 48.6K; $x$=0.25, $T_c$=40.0K; and $x$=0.40, $T_c$=31.0K. As for the $T_c$ of other metals doped cuprates, please read the relative Tables, here we omit the detail statements.

### 3.2 Oxygen content variations

The inlet picture in Figure 1 and Figure 2 show the experiment results of oxygen content concerning Fe and Al doped cuprates [41], but the results of Co doping are not illustrated because Co doped samples have the similar data to Fe and Al doped. Their oxygen content presents all the ascendant tendency in three kinds of doped cuprates. For Fe, Co and Al doping, $M^{3+}$ valence state is higher than $Cu^{2+}$ in Cu-O chains, so $M^{3+}$ ions require more oxygen ions for coordination than Cu. While the cluster do not yet form, $M^{3+}$ ions distribute randomly in individual; because the dispersed ions cannot capture one oxygen ion for charge equilibrium, therefore the oxygen content should not ascend. As $x$ increase, $M^{3+}$ ions form more and more clusters, which are able to capture oxygen ions, this is because these ions require more oxygen ions than Cu. Then the oxygen content should continuously rise in Fe, Co and Al doped cuprates, such variation characteristics are consistent with the simulated calculations. However, for Zn doping, the oxygen content falls obviously with $x$ increase [31], its variation characteristics are just opposite to the former three kinds of doped cuprates. Because Zn ions require less oxygen ions for coordination than Cu, the superfluous oxygen ions will be extruded from the clusters, therefore, oxygen vacancies would appear on $CuO_2$ planes, and then the oxygen content should descend noticeably. From the above analyses, oxygen content variation characteristics in the metal doped samples can be explained convincingly by the cluster effect, in fact, the next PAT experiments have the same conclusions.

### 3.3 Positron experiments

Concerning PAT experiments please read our other papers [28-32]. The positron lifetime is defined as the inverse of annihilation rate. According to the model of two-state capture in condensed matters [42], the short lifetime $\tau_1$ and the long lifetime $\tau_2$, the processes of positron annihilation are attributed to free-state annihilation and trapping-state annihilation, respectively. The former intensity $I_1$ denotes the proportion of free-state annihilation to total annihilation events. $\tau_1$ reflects mainly the annihilation process in the perfect crystal lattices and can be used to detect the electron distribution of the inner microstructure. And the latter intensity $I_2 = 1 - I_1$, $\tau_2$ reflects mainly the process of positron captured in the imperfect areas inside the materials, it can characterize the intrinsic structure and preparation quality of samples. If positrons annihilate in the imperfections, such as oxygen vacancy, twin boundary, dislocation, and cation vacancies, $\tau_2$ will be larger, sometimes it is designated as $\tau_3$, generally $\tau_3 > 800$ps [43-45]. Our samples do not contain $\tau_3$ component, this indicates the credibility of the sample quality.

According to the characteristics of positron annihilation, local electron density $\boldsymbol{n_e} = 1/(\pi r_o^2 c \tau_{bulk})$, where $r_o$ is the classical electron radius, $c$ is the velocity of light, and $\tau_{bulk}$ is the systematical lifetime parameter, it is defined as:

$$1/\tau_{bulk} = I_1/\tau_1 + I_2/\tau_2 . \qquad (3)$$

Some relational calculation results [46] show that about 90％ positrons annihilate with the valence electrons, and only about 10％ positrons annihilate with the electrons within the atomic kernel. These results indicate that the $\boldsymbol{n_e}$ amount is determined mainly by the valence electrons, so we may regard the density of valence electron as the $\boldsymbol{n_e}$ without influence on the general conclusions.

When $M^{3+}$ enter Cu-O chains as above statements, the dispersive $M^{3+}$ contains less one valence electron than $Cu^{2+}$, so the $\boldsymbol{n_e}$ should drop. However, the doped ions form clusters as $x$ increases, oxygen ions will be attracted into the crystal lattice. Once introduced by the clusters, every oxygen ion will catch two valence electrons in terms of charge equilibrium. Therefore the valence electron density rises, namely, the $n_e$ shows a rising trend. When the $n_e$ dropping and rising reach a balance, the low saturation emerges as listed in Table 2-4.

Obviously, such a result originate from the cluster effect. In contrast, while +2 valence ions enter $CuO_2$ planes, $Zn^{2+}$ loses fewer electrons than $Cu^{2.25+}$, so the density should rise on the consideration of valence state. However, as $x$ increase, the Zn ions begin to form clusters. As mentioned before, every four $Zn^{2+}$ will extrude an $O^-$ (including a hole) which carries away two valance electrons, so the density begin to fall. When the rising and falling of the density reach a balance, the density $n_e$ tends to high saturation, which is influenced evidently by the cluster effect. As a result, the $n_e$ saturation also can be elucidated satisfactorily through the cluster effect as the oxygen content variations above.

**Table 2. The main experimental data of $T_c$, XRD and PAT in Fe doped Y123.**

| Fe-$x$ | $T_c$(K) | a(Å) | b(Å) | $\tau_1$(ps) | $\tau_2$(ps) | $I_1$(%) |
|---|---|---|---|---|---|---|
| 0.0 | 92.0 | 3.8201 | 3.8879 | 187±3.9 | 525±22 | 87.21±1.76 |
| 0.02 | 90.0 | 3.8228 | 3.8882 | 192±3.9 | 534±22 | 86.31±1.74 |
| 0.04 | 84.0 | 3.8386 | 3.8811 | 193±3.8 | 521±22 | 84.87±1.72 |
| 0.08 | 81.1 | 3.8524 | 3.8710 | 196±3.6 | 526±22 | 86.94±1.73 |
| 0.10 | 75.3 | 3.8535 | 3.8672 | 200±3.7 | 523±22 | 85.44±1.74 |
| 0.12 | 68.0 | 3.8597 | 3.8658 | 201±3.8 | 543±22 | 86.93±1.74 |
| 0.15 | 55.5 | 3.8634 | 3.8634 | 194±3.6 | 495±22 | 84.61±1.72 |
| 0.20 | 48.6 | 3.8702 | 3.8702 | 198±3.7 | 501±21 | 85.52±1.72 |
| 0.25 | 40.0 | 3.8696 | 3.8696 | 200±3.6 | 495±22 | 85.81±1.72 |
| 0.30 |  | 3.8913 | 3.8913 | 198±3.7 | 492±22 | 84.51±1.75 |
| 0.40 | 31.0 | 3.8711 | 3.8711 | 199±3.6 | 481±22 | 84.99±1.73 |
| 0.50 |  | 3.8754 | 3.8754 | 197±3.7 | 493±22 | 84.75±1.72 |

Table 3. The main experimental data of $T_c$, XRD and PAT in Co doped Y123.

| Co-$x$ | $T_c$(K) | a(Å) | b(Å) | $\tau_1$(ps) | $\tau_2$(ps) | $I_1$(%) |
|---|---|---|---|---|---|---|
| 0.0 | 92.0 | 3.8201 | 3.8879 | 187±3.9 | 525±22 | 87.21±1.76 |
| 0.02 | 90.0 | 3.8221 | 3.8821 | 189±3.9 | 511±22 | 67.15±1.74 |
| 0.04 | 74.0 | 3.8242 | 3.8809 | 195±3.8 | 499±22 | 71.307±1.72 |
| 0.08 | 68.1 | 3.8572 | 3.8731 | 197±3.6 | 499±22 | 73.104±1.73 |
| 0.10 | 66.3 | 3.8652 | 3.8712 | 200±3.7 | 511±22 | 68.943±1.74 |
| 0.12 | 55.0 | 3.8621 | 3.8621 | 203±3.8 | 545±22 | 74.453±1.74 |
| 0.15 | 46.5 | 3.8651 | 3.8651 | 197±3.6 | 505±22 | 72.316±1.72 |
| 0.20 | 40.0 | 3.8688 | 3.8688 | 199±3.7 | 512±21 | 72.909±1.72 |
| 0.25 | 21.0 | 3.8694 | 3.8694 | 199±3.6 | 518±22 | 71.950±1.72 |
| 0.30 |  | 3.8693 | 3.8693 | 199±3.7 | 542±22 | 68.997±1.75 |
| 0.40 |  | 3.8713 | 3.8713 | 198±3.6 | 512±22 | 76.902±1.73 |
| 0.50 |  | 3.8735 | 3.8735 | 201±3.7 | 525±22 | 72.713±1.72 |

Table 4. The main experimental data of $T_c$, XRD and PAT in Al doped Y123.

| Al-$x$ | $T_c$(K) | a(Å) | b(Å) | $\tau_1$(ps) | $\tau_2$(ps) | $I_1$(%) |
|---|---|---|---|---|---|---|
| 0.0 | 92.0 | 3.8155 | 3.866 | 195±3.9 | 619±22 | 94.12±1.77 |
| 0.05 | 91.5 | 3.8252 | 3.8621 | 196±3.8 | 581±22 | 94.04±1.76 |
| 0.1 | 91.5 | 3.8408 | 3.8553 | 197±3.9 | 553±22 | 92.21±1.77 |
| 0.15 | 87 | 3.8492 | 3.8486 | 203±3.8 | 552±22 | 91.87±1.75 |
| 0.2 | 83.2 | 3.8485 | 3.8485 | 206±3.6 | 549±21 | 92.10±1.73 |
| 0.3 | 72.5 | 3.8511 | 3.8511 | 201±3.7 | 565±22 | 93.32±1.76 |
| 0.4 | 42.5 | 3.8543 | 3.8543 | 205±3.6 | 557±22 | 92.84±1.74 |

## 3.4 Valence electron density and the $T_c$ of cuprates

Figure 3 shows the relationship between the $T_c$ and the reduced $n_e$ in Fe, Co and Al doped cuprates, for the sake of comparison with Zn doped cuprates, Figure 3 illustrates yet the correlated data of Zn doping.

For Fe doped samples, the reduced $n_e$ has the great decrease in the little doping, the $T_c$ descends only slightly as shown in Figure 3, however, the $n_e$-$T_c$ curve shows a turning point as Fe concentration $x$ increase. Below the curve turning point, the $T_c$ seems to lose further the association with the reduced density $n_e$ or the valence electrons. And Co doped samples are similar to Fe doped in the relationship between the $T_c$ and the reduced $n_e$, here we omit the detail statements. As for Al doped cuprates, in the little doping, the relationship between the $T_c$ and $n_e$ is the same as Fe and Co, namely, the reduced $n_e$ shows the great decrease when the $T_c$ descends only slightly. But the reduced $n_e$ of Al doping descends slower than that of Fe and Co doped samples in the same concentration $x$, as shown in Figure 3, so the $n_e$-$T_c$ curve turning point in Fe and Co doped samples are further from $n_e =1.00$ than that in Al doped, such a result is yet evidence for that the $E_m$ is larger in Fe, Co doped samples. Anyway, in three kinds of doped cuprates, below the curve turning point the $T_c$ seems to lose further the association with the reduced density $n_e$ or the valence electrons, in particular, when their $T_c$ drop sharply, the reduced $n_e$ only slightly vary as shown in Figure 3.

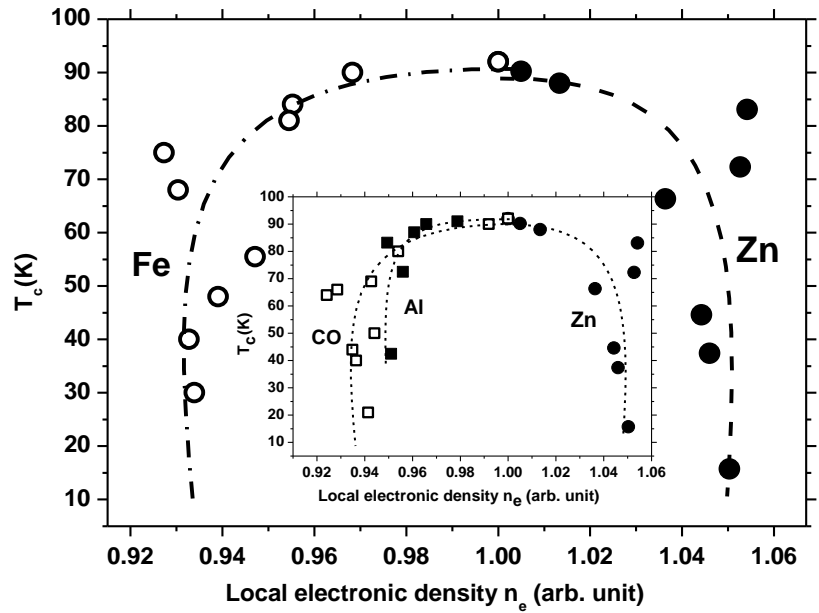

**Figure 3.** The variation of superconducting transition temperature Tc with the reduced local electron density ne in YBa2Cu3-x(Fe, Zn)xO7-δ cuprates. For the sake of illustration effect, the inlet picture indicates the experimental results in YBa2Cu3-x(Co, Al, Zn)xO7-δ cuprates.

In fact, when $x$ is a little, $M^{3+}$ ions enter the crystal lattice in individual, they loses more valence electrons than $Cu^{2+}$, so the density $n_e$ will decrease. However, the $T_c$ descends slightly with the density change, because the dispersed ions may not destroy the crystal lattice. Below the curve turning point the $T_c$ starts to drop sharply with clusters appearing, on our idea, the cluster effect distorts the crystal texture of cuprates, so the superconductivity is suppressed seriously. Therefore, the cluster effect is an important factor that suppresses the cuprate superconductivity, the same conclusion can be achieved from the following discussions concerning the Zn doped samples. In contrast to $M^{3+}$ doping above, the $n_e$-$T_c$ curve of Zn doping has the opposite variation trends in the beginning stage, though the reduced density raises greatly, the $T_c$ descend still slightly, below the curve turning point the $T_c$ starts to drop sharply from 72.3K to 15.7K, the reduced $n_e$ only slightly increase. As above statements, similarly, when $x$ is a little, Zn ions enter the crystal lattice in individual; $Zn^{2+}$ loses fewer electrons than $Cu^{2.25+}$, so the density will increase. However, the $T_c$ descends slightly with the density increasing, because the dispersed ions may not destroy the crystal lattice. While the $Zn^{2+}$ clusters appear in the doped samples, the crystal texture will be distorted by the cluster effect, which will influence directly the pairing and transportation of carriers, and thus the superconductivity is suppressed markedly, then the $T_c$ starts to drop sharply. On all accounts, when the reduced density increases or decreases evidently, the $T_c$ only alters slightly; contrary, while the $T_c$ falls dramatically, the reduced density hardly changes. As a result, we can conclude that the $T_c$ has no direct relationship with the valence electron density or the valence electrons.

## 4. Conclusions

The high-$T_c$ cuprates $YBa_2Cu_{3-x}$(Fe, Co, Al)$_x$O$_{7-\delta}$(x=0.0～0.5) have been analyzed and studied systemically by means of XRD, positron annihilation technique, oxygen content and calculations of binding energy, for the sake of comparison with Zn doped cuprates, we mentions yet some correlated contents of the latter. In conclusion, the experimental results and simulated calculations support that the $T_c$ does not depend directly on the density of valence electron in the samples. At the same time, on our idea from the investigations, whether or not it may imply some relations between Fe-oxide superconductors and cuprates that Fe doped Y123 samples show some unique characteristics.